# Relativistic laser driven electron accelerator using micro-channel plasma targets

J. Snyder[1+], L. L. Ji[2,3*], K. M. George[4], C. Willis[5], G. E. Cochran[5], R. L. Daskalova[5], A. Handler[6], T. Rubin[7], P. L. Poole[6], D. Nasir[5], A. Zingale[5], E. Chowdhury[5], B. F. Shen[2,3,8], D. W. Schumacher[5]

[1]*Department of Mathematical and Physical Sciences, Miami University, Hamilton, OH 45011, USA*
[2]*State Key Laboratory of High Field Laser Physics, Shanghai Institute of Optics and Fine Mechanics, Chinese Academy of Sciences, Shanghai 201800, China*
[3]*CAS Center for Excellence in Ultra-intense Laser Science, Shanghai 201800, China*
[4]*Innovative Scientific Solutions Inc., Dayton, OH 45459*
[5]*Department of Physics, The Ohio State University, Columbus, OH 43210, USA*
[6]*Lawrence Livermore National Laboratory, Livermore, CA 94550, USA*
[7]*Voss Scientific, Albuquerque, NM 87108, USA*
[8]*Shanghai Normal University, Shanghai 200234, China*

**Abstract**

We present an experimental demonstration of the efficient acceleration of electrons beyond 60 MeV using micro-channel plasma targets. We employed a high-contrast, 2.5 J, 32 fs short pulse laser interacting with a 5 μm inner diameter, 300 μm long micro-channel plasma target.  The micro-channel was aligned to be collinear with the incident laser pulse, confining the majority of the laser energy within the channel. The measured electron spectrum showed a large increase of the cut-off energy and slope temperature when compared to that from a 2 μm flat Copper target, with the cutoff energy enhanced by over 2.6 times and the total energy in electrons >5 MeV enhanced by over 10 times. Three-dimensional particle-in-cell simulations confirm efficient direct laser acceleration enabled by the novel structure as the dominant acceleration mechanism for the high energy electrons. The simulations further reveal the guiding effect of the channel that successfully explains preferential acceleration on the laser/channel axis observed in experiments. Finally, systematic simulations provide scalings for the energy and charge of the electron pulses. Our results show that the micro-channel plasma target is a promising electron source for applications such as ion acceleration, Bremsstrahlung X-ray radiation, and THZ generation.

*jill@siom.ac.cn

+snyderjc@miamioh.edu

# I. Introduction

The relativistic laser-plasma interaction (LPI) is an efficient source of high energy electrons [1-5], ions [6-8], high-order harmonics [9, 10] and electron-positron jets [11-13]. These processes mostly rely on energy conversion from the laser pulse to electrons, since the latter is more responsive to the laser field than other species. In the low plasma-density regime, both direct laser acceleration (DLA) and laser-driven wakefield acceleration (LWFA) are possible mechanisms for generating high energy electrons. Electrons gain energy directly from the oscillating laser field in the former case and from the plasma wakefield in the latter. Steady progress has been made with low-density gas targets, from which multi-GeV electrons are now available [14, 15]. The interaction at low densities tends to be controllable wherein one can tune the energy, charge, and energy spread of the electron beam by varying the density, injection position, and composition of the target [16-19]. The generation of superponderomotive electrons in relativistically transparent and near-critical density plasmas has been studied experimentally and via simulations, although the propagation length of the laser pulse through these media may be limited [20-26]. The scenario in the high density regime, however, is quite different. A highly overdense plasma interface prohibits a short laser pulse from propagating through unless it is ultrathin, thus the interaction region of the laser field and a given electron is usually restricted to a length of order the laser wavelength. The energy gain for electrons is therefore limited, leading to a relatively low energy/efficiency that scales with the laser intensity as $\sim \sqrt{I_0}$ [27]. Moreover, the investigation of ultra-short lasers interacting with initially solid-density matter has been mainly focused on flat targets, where controlling LPI is more challenging partially due to the pre-pulse issue [28].

A recent development to enhance the laser-solid interaction is using structured interfaces [29-45]. Particularly, nano-wires [32], nano-particles [29], nano-spheres [31] and snowflakes [30] have been proposed to increase the laser absorption efficiency. These structures are mostly on a nanometer scale that promotes efficient surface heating of electrons and hence increases their population. An alternative approach with the potential for a larger boost in electron energies uses laser pulses and target structures designed for a prolonged interaction. For a laser wavelength ~1 μm, we suggest that structures on a comparable scale offer complementary capabilities. They can enable the controlled propagation of the laser beam, turning the DLA mechanism on in the solid density regime. DLA is superior to surface heating for producing directed, high-energy electrons [46, 47]. Its possible advantages for enhancing proton acceleration [34, 39] and local laser intensities [38] have been shown in previous simulations. We therefore propose to use micro-sized structures to extend the laser-solid interaction in a highly controlled fashion. Practical application requires that the micro-scale structures be produced in a well aligned and highly-ordered array. The ordered structures can be altered to vary the spatial period, spacing and/or length of the structures to tailor the laser-driven electron source to a specific desired secondary application, such as ion acceleration, Bremsstrahlung X-ray radiation, and THZ generation.

In this paper, we present experimental results using a micro-channel structure to enhance and manipulate laser-driven electron sources via DLA in the solid density regime. Using the ultrahigh intensity Scarlet Laser [48] at The Ohio State University, we demonstrate a significant increase in cut-off energy and electron yield using the micro-channel structure compared to a flat, 2μm Cu target. Additionally, we note a preferential acceleration along the tube axis by comparing electron spectra along both the tube axis and the target normal axis. Three-dimensional particle-in-cell (PIC) simulations confirm the role of DLA when using the micro-channel targets and elucidate the guiding capabilities of the novel target structure.

## II. Results

### A. Experiment

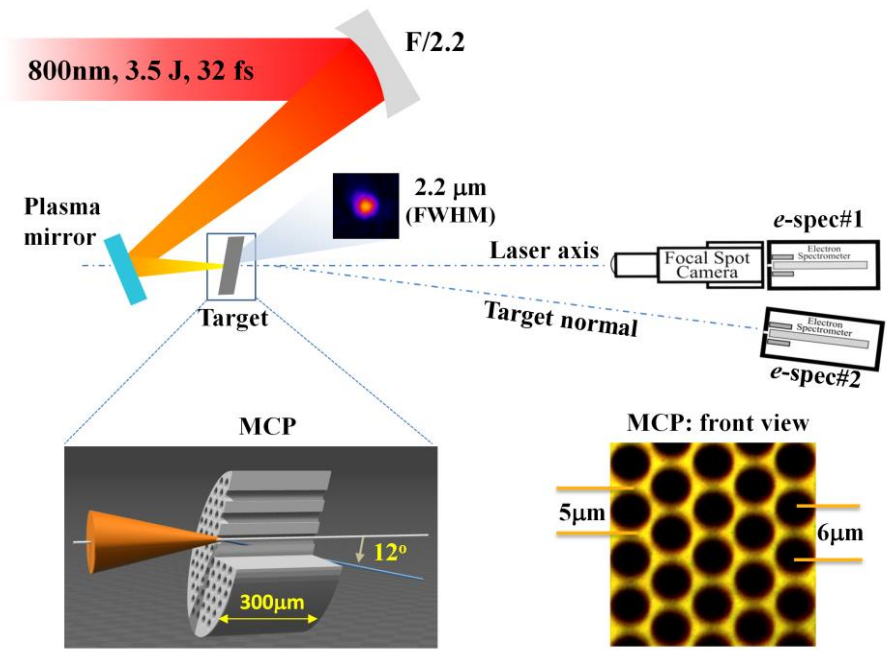

**Figure 1.** Experimental set-up: An 800 nm, 3.5 J, 32 fs laser pulse is focused onto a plasma mirror, resulting in ~2.5 J delivered onto the target in a focal spot of 2.2 μm full-width at half-maximum. The 300 μm long channels of the MCP (inset) are aligned

with the laser propagation axis. The laser is incident on the target surface at the channel 12° bias angle with respect to the surface. The electron signal is measured with magnetic spectrometers on the laser/channel axis (e-spec #1) and along the rear target normal direction (e-spec #2). Lower-right image shows the front view of MCP channel with an inner diameter of 5 μm and a spatial period is 6 μm.

The experimental setup is detailed in Figure 1. During the experiment, the Scarlet laser delivered an 800 nm-wavelength, 32 fs-duration beam with total energy averaging 3.5 J. An F/2.2 off-axis parabola was used to focus the beam. The intrinsic pulse contrast was greater than $10^{10}$ on the nanosecond scale and $10^8$ on the picosecond scale. The contrast was further enhanced by 30-40 times by placing a plasma mirror (PM) 2.8 mm before the target. After the PM, 2.5 J of laser energy was left, 80% of which was focused onto a spot size as small as 2.2 μm full-width at half-maximum (FWHM), resulting in an on-target intensity up to $6.7 \times 10^{20}$ W/cm$^2$. Considering misalignment between the laser and the target, the estimated on-target intensity could be lower. Comparison of electron spectra from flat foil targets with simulations suggest a peak intensity around $4.2 \times 10^{20}$ W/cm$^2$.

We chose a micro-channel plate (MCP) produced by Photonis USA, Inc. as the structured target. The MCP was 300 μm thick, with a channel inner diameter of 5 μm, pitch of 6 μm, and bias angle of 12°. The 12° bias angle allowed target alignment with the tube axis parallel to the laser axis without having back reflections that could possibly damage optics in the laser chain. The 6 μm pitch provided a favorable open area ratio that ensured most of the laser energy entered the structure while the 5 μm inner diameter ensured the laser intensity at the boundary would be sufficient to enhance the interaction. Multiple targets were cut from a single device. These features made the MCP target an ideal, and readily available, structure for manipulating DLA in the solid-density regime. MCPs are designed for measuring electron/X-ray signals wherein a voltage across the MCP produces cascading electrons that enhance the input signal. They are relatively low cost and readily available and here we show that the unique properties of the MCP are most favorable for enhancing and manipulating LPI in the overdense regime.

To characterize the angular distribution, accelerated electrons were collected in two magnetic spectrometers located along the laser-axis (or channel axis, defined as 0º) and along the target normal (12º off laser-axis). The entrance slit of both spectrometers was 0.5 cm high and 250 μm wide. The 0º spectrometer had an energy range of 0.3-61 MeV while the energy range for the 12º spectrometer was 0.3-63 MeV. The magnetic field in the center of the gap of the 0º and 12º spectrometers measured 0.55 T and 0.58 T, respectively. The spectrometers used BAS-MS image plate (IP) detectors scanned using a GE Typhoon FLA 7000 scanner. The raw signal of the image plates (Figure 2) was converted to total electron number using the method described by Tanaka, et al. [49] with the IP calibration given by Rabhi, et al. [50] For comparison shots, we used a standard 2 μm flat Cu target and a 300 μm flat glass target. The latter was cut from the edge of the MCP where no channels were present to have a material matching that of the channel walls.

A significant difference can be seen from the raw signals received by the IPs in Figure 2(a-b). The signal from the MCP target is saturated on the scale shown, while that from the (unstructured) 2 μm Cu foil is much weaker. The derived electron energy spectra are compared in Figure 2. The electron cutoff energy from the MCP target extends beyond the detectable range of either spectrometer, reaching a maximum energy in excess of 63 MeV – at least 2 times higher than the Cu foil cutoff. When compared to the baseline 300 μm glass target, there is nearly an order of magnitude increase in cutoff energy for the MCP target. Further, the electron slope temperature for the MCP targets reaches up to 15 MeV, much higher than the flat interfaces in either direction. The total energy in electrons >5 MeV from the MCP is enhanced by more than an order of magnitude when compared to the 2 μm Cu target.

It has been known that an ultra-intense laser normally impinging on a solid flat foil creates a distribution with an approximate, characteristic ponderomotive electron temperature of $kT_e \text{ [MeV]} \approx 0.511 \times (\sqrt{1 + I_0 \lambda_0^2 / 1.37 \times 10^{18} \text{W} \mu\text{m}^2/\text{cm}^2} - 1)$ [27], with $I_0$ the peak intensity in W/cm$^2$ and the wavelength $\lambda_0$ in μm. For parameters specified above, the slope temperature is predicted to be ~6.7 MeV, consistent with our results for the flat Cu-target (see in Figure 3). However, the MCP target produces superponderomotive electrons with an electron slope temperature that is more than twice the predicted value, suggesting a different acceleration mechanism.

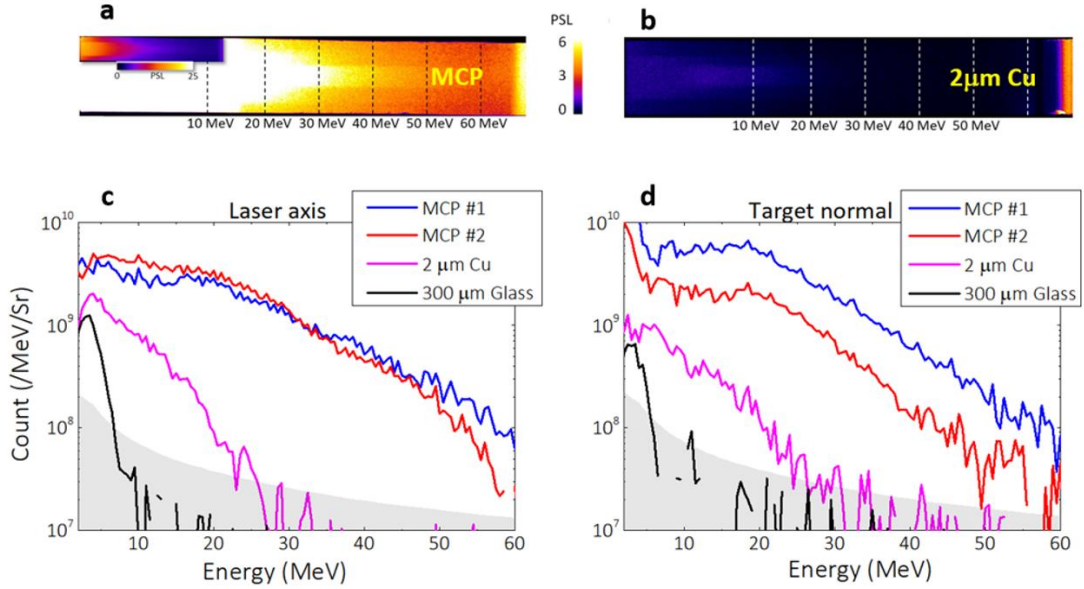

**Figure 2**. Raw image plate signal from (a) the MCP target and (b) the 2μm Cu target on the same scale. The inset in (a) shows the MCP target signal on a different scale. The derived electron spectra on (c) laser/channel axis and (d) target normal. Shown are results using MCP targets (blue, red), the 2 μm thick Cu targets (pink), and the 300 μm thick glass target (black). The gray area denotes the detection limits of the electron-spectrometers.

The slope temperature for each case is summarized in Figure 3. One notices the distinctive trend between the two measured directions for each target type. The trend seen in the Cu foil is typical for simple flat foils where the laser does not break through. For such a case, the sheath field set up at the rear surface points in the target normal direction. The majority of the super-thermal electrons are deflected by the sheath field each time they circulate though the foil [51] and finally, on average, these electrons expand along the target normal direction. This is clearly illustrated in Figure 3, where the target-normal slope temperature is larger than the on-axis one for the flat target.

This is in contrast to the MCP target which produces higher slope temperatures on the laser/channel axis than in the other direction, as shown in Figure 3. This suggests a preferential acceleration of the highest energy electrons along the $0^0$ direction. As we discuss below, this is strong evidence of the guiding capability of the MCP targets.

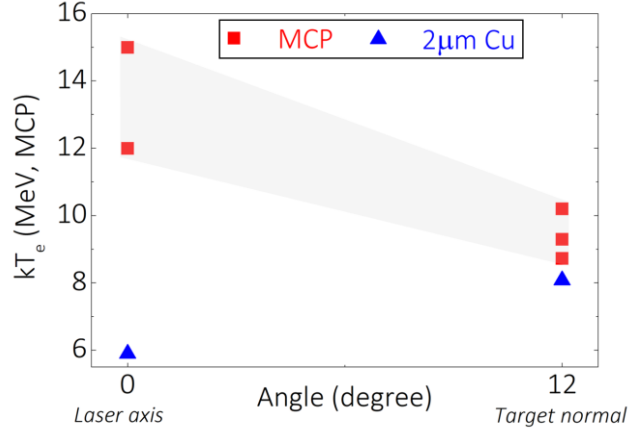

**Figure 3**. The slope temperatures from MCP targets (red-squares) and 2μm Cu targets (blue-triangles) in the experiment. The data was collected on the laser axis at 0º angle and in the target normal direction at 12º angle.

### B. Simulations

The above experimental results show that the MCP structures greatly enhance both the maximum energy and the generation efficiency of the laser-driven electrons in the solid-density regime. To establish the acceleration mechanism, we carried out full three-dimensional (3D) particle-in-cell (PIC) simulations with the simulation code VLPL [52]. In each simulation, a $y$-polarized laser pulse propagates along the $x$ direction in a simulation box of $40\lambda_0 \times 22\lambda_0 \times 22\lambda_0$ extent divided into $1600\times220\times220$ cells in the $x \times y \times z$ dimensions, respectively ($\lambda_0$= 0.8μm). The laser field amplitude has a profile of $a_y = a_0 e^{-(r^2/\sigma_0^2)} \sin^2[\pi t/(2\tau_0)]$ where $a_0 = eE_0/m_e\omega_0 c$ is the dimensionless laser electric field amplitude with $e$ and $m_e$ the fundamental charge and electron mass, $E_0$ and $\omega_0$ the amplitude and angular frequency of the laser, and $c$ the light speed in vacuum. We set $a_0 = 14$ ($I_0 \approx 4.2\times10^{20}\,\text{W/cm}^2$), pulse duration $\tau_0 = 35$ fs and spot size $\sigma_0 = 2.56$ μm, corresponding to a total energy of 2.15 J. To mimic the laser interacting with a single channel, a 300 μm long carbon tube with inner diameter of 5 μm and boundary thickness of 1 μm was placed at $x = 8$ μm. A moving window was employed to model the long interaction distance. The tube is fully ionized with an initial electron density of $n_e = 180 n_c$ ($n_c = m_e \omega_0^2 / 4\pi e^2$ is the critical density). The time step was $\Delta t = 0.008 T_0$ to meet the criterion for relativistic electron motion [53]. We used 27 macro-particles per cell for all species.

The results for a 2 μm thick Cu-foil and the MCP target are shown in Figure 4 (a). The difference between the MCP target and the flat target is consistent with the experimental results. The slope temperature from the forward-moving electrons is around 16.5 MeV for the MCP simulation and 7.5 MeV for the Cu-foil, in excellent agreement with our experimental observations.

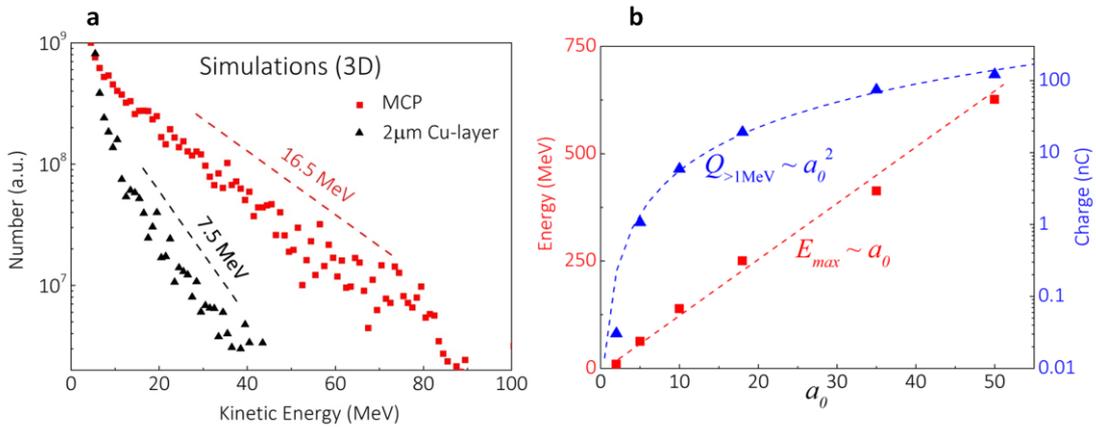

**Figure 4**. (a) Electron energy spectra from 3D PIC simulations for laser interaction with the MCP target (red-squared) and 2 μm Cu-layer (black-triangled). Parameters were chosen to match experimental condition ($a_0 = 14$, tube length 300 μm, inner diameter 5 μm etc.). The simulation spectra were recorded at the end of the simulation. (b) Scaling of the cut-off electron energy and total charge (>1MeV) as a function of the laser amplitude from simulations. The dashed lines are fits of the simulation data. Simulation data were collected when the acceleration peaks.

## III. Discussion

We now consider the potential for this approach as an electron source by studying its scaling properties using a series of 3D PIC simulations. As shown in Figure 4 (b), the resulting cut-off energy and total beam charge (> 1MeV) were presented for laser amplitudes up to $a_0$ = 50. We derived the scaling behavior by fitting the simulated results with a linear fit for cut-off energy and a parabolic fit for the beam charge. The cut-off electron energy is found to be proportional to the laser amplitude. While one would expect a scaling ~ $a_0^2$ for DLA in free space, the weaker scaling found here could be partially due to the superluminal phase velocity in a channel [54]. Also, we find that the laser field extracts a significant number of electrons from the boundary that compensate the laser field so that the accelerating field is weakened. This "overloading" effect has also been noted in wakefield acceleration [55] and is more evident for stronger laser fields, leading to weaker energy scaling.

On the other hand, efficient loading of electrons into the laser field yields $\sim a_0^2$ scaling of the beam charge. Electrons extracted from the channel extend to a transverse length of $l_e \sim \beta_\perp \Delta t \sim \beta_\perp / (1 - \beta_x) = \sqrt{(1 + \beta_x)/(1 - \beta_x)}$ ($\beta_\perp$ and $\beta_x$ are the transverse and longitudinal velocities of the electrons; $\Delta t \sim 1/(1-\beta_x)$ is the dephasing time between the electrons and the laser pulse). The in-tube electron density $n_e \sim a_0$ [37], thus the total electron number scales as $N_e \sim n_e l_e \sim a_0 p_x$ for $p_x \gg 1$. The scaling law $p_x \sim a_0$ yields $N_e \sim a_0^2$ which agrees with the results shown in Fig. 4(b). With currently available laser technology ($a_0 = 50$, beam energy 20 J), electron pulses with a total charge over 100 nC at $kT_e \sim 125 \text{MeV}$ (23% efficiency) can be obtained. Therefore the laser-driven micro-channel-array could be a favorable electron source for numerous applications such as ion acceleration, Bremsstrahlung X-ray source, electron-positron jets, and intense THz generation, among others.

It should be mentioned that electron acceleration within the channel could reach the dephasing stage if the tube is too long. We find that the cut-off energy for a laser pulse with $a_0 = 50$ maximizes at $t = 50T_0$, corresponding to a ~35 μm acceleration length. Detailed studies indicate the dephasing length for $a_0 = 5$-50 lies between 20-50 μm, although the underlying mechanism requires further investigation. The simulation results in Figure 4(a) were obtained after a 300 μm interaction distance to mimic the experimental process, while in Figure 4(b) the electron spectrum data were collected at the time the acceleration peaks. Hence the cut-off energy is higher in the latter at $a_0 = 14$.

An obvious advantage of the MCP target is that the driving laser can propagate in the channel for a distance much larger than the Rayleigh length. Hence the laser intensity can be preserved for a long time. In our case, the pulse was able to transmit through the 300 μm long channel before losing 95% of its total energy. During propagation, the transverse laser electric field is strong enough to extract electrons directly from the inner channel surface. As shown in Figure 5(a), a positive $E_y$ field detaches electrons from the upper boundary ($y>0$) and pulls them towards the center of the tube, while a negative $E_y$ field pulls electrons from the lower boundary towards the center. These electrons gain relativistic velocities in the transverse direction in less than half of a laser period. Then, the laser $B_z$ field turns the electrons in the forward direction, via the direct laser acceleration mechanism. Although the field exerted on these electrons is relatively weak at the boundary, the electrons are moved towards the channel center where the field peaks. This leads to a large energy gain for a significant portion of the electrons. In the end, a train of micro-bunches is generated with energies extending beyond 100 MeV in the simulations.

In free space, direct acceleration of electrons by a propagating laser is subject to the transverse ponderomotive force. Electrons injected head on into the laser field can be scattered, resulting in limited energy gain and large divergence angle. Instead, when using a channel, the electrons are injected transversely and the channel itself acts as a guiding structure for the energetic electrons (as shown below) and the laser pulse. This enables a long interaction distance. In Figure 5(b), we show the laser-induced plasma field in the vicinity of the tube. The plasma quasi-static fields are obtained by averaging the fields over one pulse length (FWHM) in the $x$ direction for a given time step. This approximately eliminates the oscillating laser field. We observe a charge separation field $E_p$ created by the electrons extracted from the channel walls and a surrounding magnetic field $B_p$. The latter is from the return current within the channel walls that compensates for the current caused by the forward going electrons in the channel. While the $E_p$ field tends to pull the electrons back to the boundary, the $B_p$ field pushes them towards the center. The E-field and B-field ultimately have the same source, the extracted relativistic electrons, and thus are coupled together yielding forces that are approximately equal in magnitude. Hence, electrons can flow along the channel without diverging as they do in free space.

To describe this, we calculated the electron energy density distribution in angular space. Each electron in the channel is characterized by a divergence angle $\theta$ between the electron momentum and the propagation axis $x$ and a lateral angle $\varphi$ between the transverse momentum $p_\perp = \sqrt{p_y^2 + p_z^2}$ and the $y$-axis. In Figure 5(c) and (d), the angular distribution is shown for the MCP and the flat Cu-target. The laser/channel axis corresponds to $\theta = 0°$ while the target normal direction is at ($\theta = 12°$, $\varphi = 180°$). Electrons with kinetic energy exceeding 1 MeV emitted from the flat Cu-target are peaked at $\theta = 12°$ and $\varphi = 180°$ - exactly the target normal direction. In contrast, for the MCP target the majority of the high energy electrons are centered at $\theta = 0°$. The result is consistent with our experimental observation in Figure 3, providing strong evidence of the channel's guiding capabilities through the entire length of the channel. We note that electrons from the Cu-target are generated within a divergence angle of $\theta \approx 11°$ (half width at half maximum). As a comparison, a well-collimated electron beam is produced within a divergence angle $\theta \approx 3.5°$ for the MCP target, a dramatic improvement on the beam quality due to the guiding effect.

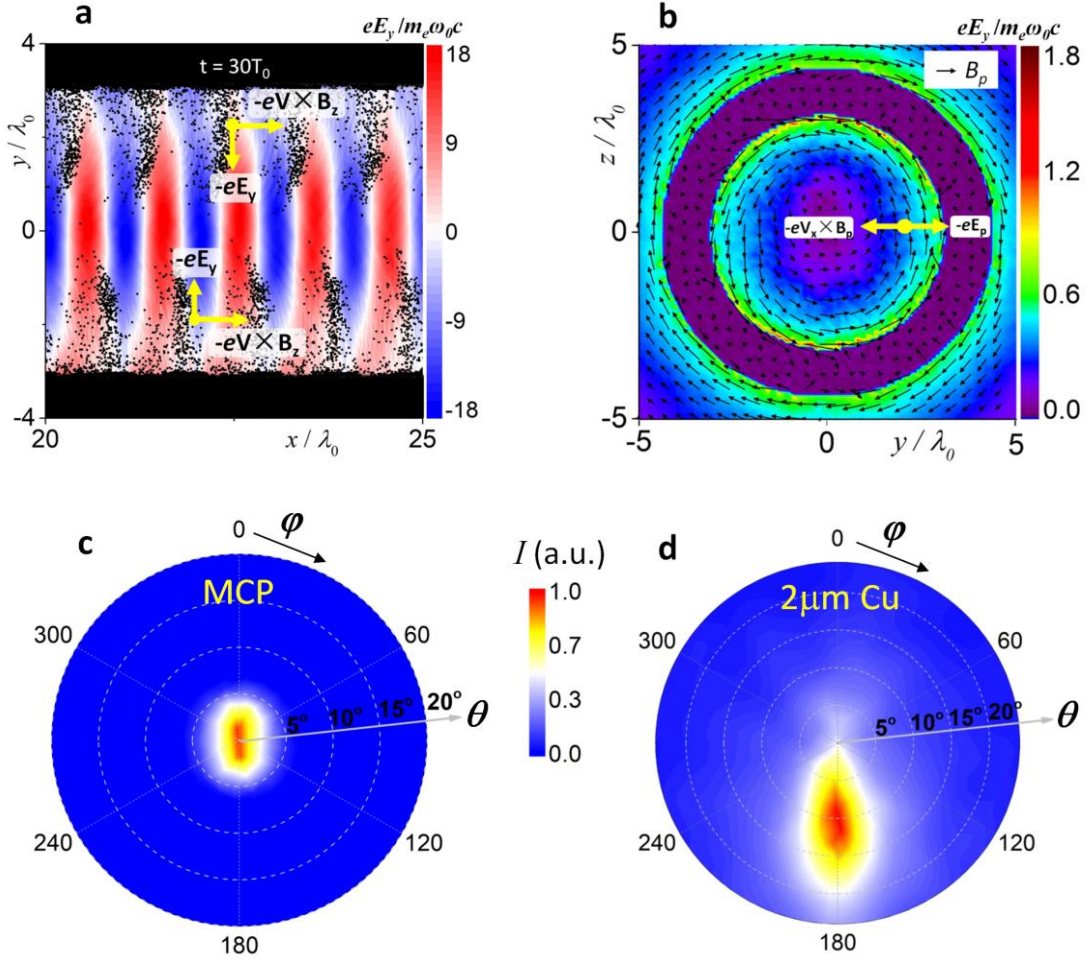

**Figure 5**. (a) Distributions of the electron density and laser field $E_y$ within the channel in the *x-y* plane ($z=0$). (b) The averaged charge separation field $E_p$ and angular magnetic field $B_p$ in the vicinity of the channel. The angular distribution of the electron energy density for the MCP (c) and the Cu target (d).

## IV. Conclusions

In conclusion, we have proposed and successfully demonstrated high-energy electron acceleration produced by laser-driven micro-channel plasma targets with both experiments and simulations. The efficient DLA enabled by the novel micro-structure greatly enhances the outcome of the LPI and makes manipulation of the interaction possible. Secondary applications based on the micro-structures are currently under investigation.

## V. Acknowledgements

This work is supported by the AFOSR Basic Research Initiative (BRI) under contract FA9550-14-1-0085 and allocations of computing time from the Ohio Supercomputing Center. LJ acknowledges the National Science Foundation of China (Nos. 11374317 and 11335013), the Strategic Priority Research Program of Chinese Academy of Sciences (Grant No. XDB16000000) and the Recruitment Program for Young Professionals, for support. GEC, PLP, AZ, and DWS acknowledge the U.S. Department of Energy, National Nuclear Security Administration, under contract DE-NA0003107.

**References**

[1] J. Faure, Y. Glinec, A. Pukhov, S. Kiselev, S. Gordienko, E. Lefebvre, J.P. Rousseau, F. Burgy, and V. Malka, Nature 431, 541 (2004).

[2] C. G. R. Geddes, C. Toth, J. Van Tilborg, E. Esarey, C. B. Schroeder, D. Bruhwiler, C. Nieter, J. Cary, and W. P. Leemans, Nature 431, 538 (2004).

[3] S. P. D. Mangles, C. D. Murphy, Z. Najmudin, A. G. R. Thomas, J. L. Collier, A. E. Dangor, E. J. Divall, P. S. Foster, J. G. Gallacher, C. J. Hooker, D. A. Jaroszynski, A. J. Langley, W. B. Mori, P. A. Norreys, F. S. Tsung, R. Viskup, B. R. Walton, and K. Krushelnick Nature 431, 535 (2004).

[4] K. B. Wharton, S. P. Hatchett, S. C. Wilks, M. H. Key, J. D. Moody, V. Yanovsky, A. A. Offenberger, B. A. Hammel, M. D. Perry, and C. Joshi, Physical review letters 81, 822 (1998).

[5] G. Malka and J. L. Miquel, Physical review letters 77, 75 (1996).

[6] R. A. Snavely, M. H. Key, S. P. Hatchett, T. E. Cowan, M. Roth, T. W. Phillips, M. A. Stoyer, E. A. Henry, T. C. Sangster, M. S. Singh, S. C. Wilks, A. MacKinnon, A. Offenberger, D. M. Pennington, K. Yasuike, A. B. Langdon, B. F. Lasinski, J. Johnson, M. D. Perry, and E. M. Campbell, Physical review letters 85, 2945 (2000).

[7] B. M. Hegelich, B. J. Albright, J. Cobble, K. Flippo, S. Letzring, M. Paffett, H. Ruhl, Jörg Schreiber, R. K. Schulze, and J. C. Fernández, Nature 439.7075, 441 (2006).

[8] H. Habara, K. L. Lancaster, S. Karsch, C. D. Murphy, P. A. Norreys, R. G. Evans, M. Borghesi, L. Romagnani, M. Zepf, T. Norimatsu, Y. Toyama, R. Kodama, J. A.

King, R. Snavely, K. Akli, B. Zhang, R. Freeman, S. Hatchett, A. J. MacKinnon, P. Patel, M. H. Key, C. Stoeckl, R. A. Fonseca, and L. O. Silva, Physical Review E 70, 46414 (2004).

[9] B. Dromey, M. Zepf, A. Gopal, K. Lancaster, M. S. Wei, K. Krushelnick, M. Tatarakis, N. Vakakis, S. Moustaizis, R. Kodama, M. Tampo, C. Stoeckl, R. Clarke, H. Habara, D. Neely, S. Karsch, and P. Norreys, Nature physics 3, 456-459 (2006).

[10] C. Thaury and F. Quéré, Molecular and Optical Physics 43, 213001, (2010).

[11] T. E. Cowan, M. D. Perry, M.H. Key, T. R. Ditmire, S. P. Hatchett, E. A. Henry, J. D. Moody, M. J. Moran, D. M. Pennington, T. W. Phillips, T. C. Sangster, J. A. Sefcik, M. S. Singh, R. A. Snavely, M. A. Stoyer, S. C. Wilks, P. E. Young, Y. Takahashi, B. Dong, W. Fountain, T. Parnell, J. Johnson, A. W. Hunt, and T. Kuhl, Laser and particle beams 17, 773 (1999).

[12] C. Gahn, G. D. Tsakiris, G. Pretzler, K. J. Witte, C. Delfin, C-G. Wahlström, and D. Habs, Applied physics letters 77, 2662 (2000).

[13] H. Chen, S. C. Wilks, J. D. Bonlie, E. P. Liang, J. Myatt, D. F. Price, D. D. Meyerhofer, and P. Beiersdorfer, Physical review letters 102, 105001 (2009).

[14] W. P. Leemans, B. Nagler, A. J. Gonsalves, C. Toth, K. Nakamura, C. G. Geddes, E. Esarey, C. B. Schroeder, and S. M. Hooker, Nature physics 2, 696-699 (2006).

[15] X. Wang, R. Zgadzaj, N. Fazel, Z. Li, S. A. Yi, X. Zhang, W. Henderson, Y. Y. Chang, R. Korzekwa, H. E. Tsai, C. H. Pai, H. Quevedo, G. Dyer, E. Gaul, M. Martinez, A. C. Bernstein, T. Borger, M. Spinks, M. Donovan, V. Khudik, T. Ditmire, and M. C. Downer, Nature communications 4, 1988 (2013).

[16] W. T. Wang, W. T. Li, J. S. Liu, Z. J. Zhang, R. Qi, C. H. Yu, J. Q. Liu, M. Fang, Z. Y. Qin, C. Wang, Y. Xu, F. X. Wu, Y. X. Leng, R. X. Li, and Z. Z. Xu, Physical review letters 117, 124801(2016)

[17] C. Rechatin, J. Faure, A. Ben-Ismail, J. Lim, R. Fitour, A. Specka, H. Videau, A. Tafzi, F. Burgy, and V. Malka, Physical review letters 102, 164801 (2009).

[18] M. Mirzaie, S. Li, M. Zeng, N. A. M. Hafz, M. Chen, G. Y. Li, Q. J. Zhu, H. Liao, T. Sokollik, F. Liu, Y. Y. Ma, L. M. Chen, Z. M. Sheng, and J. Zhang, Scientific reports 5, 14659 (2015).

[19] B. B. Pollock, C. E. Clayton, J. E. Ralph, F. Albert, A. Davidson, L. Divol, C. Filip, S. H. Glenzer, K. Herpoldt, W. Lu, K. A. Marsh, J. Meinecke, W. B. Mori, A. Pak, T. C. Rensink, J. S. Ross, J. Shaw, G. R. Tynan, C. Joshi, and D. H. Froula, Physical review letters 107, 045001 (2011).

[20] L. Willingale, A. V. Arefiev, G. J. Williams, H. Chen, F. Dollar, A. U. Hazi, A. Maksimchuk, M. J. Manuel, E. Marley, W. Nazarov, T. Z. Zhao, and C. Zulick, New journal of physics 20, 093024 (2018).

[21] O. N. Rosmej, N. E. Andreev, S. Zaehter, N. Zahn, P. Christ, B. Borm, T. Radon, A. Soko-lov, L. P. Pugachev, D. Khaghani, F. Horst, N. G. Borisenko, G. Sklizkov, and V. G. Pimenov, arXiv preprint arXiv:1811.01278 (2018).

[22] J. H. Bin, M. Yeung, Z. Gong, H. Y. Wang, C. Kreuzer, M. L. Zhou, M. J. V. Streeter, P. S. Foster, S. Cousens, B. Dromey, J. Meyer-ter-Vehn, M. Zepf, and J. Schreiber, Physical review letters 120, 074801 (2018).

[23] D. J. Stark, T. Toncian, and A. V. Arefiev, Physical review letters 116, 185003

(2016).

[24] T. W. Huang, C. T. Zhou, H. Zhang, S. Z. Wu, B. Qiao, X. T. He, and S. C. Ruan, Applied physics letters 110, 021102 (2017).

[25] M. Vranic, R. A. Fonseca, and L. O. Silva, Plasma physics and controlled fusion 60, 034002 (2018).

[26] S. Feister, D. R. Austin, J. T. Morrison, K. D. Frische, C. Orban, G. Ngirmang, A. Handler, J. R. Smith, M. Schillaci, J. A. LaVerne, E. A. Chowdhury, R. R. Freeman, and W. M. Roquemore, Optics express 25, 18736 (2017).

[27] S. C. Wilks, W. L. Kruer, M. Tabak, and A. B. Langdon, Physical review letters 69, 1383 (1992).

[28] V. V. Ivanov, A. Maksimchuk, and G. Mourou, Applied optics 42, 7231 (2003).

[29] P. P. Rajeev, P. Ayyub, S. Bagchi, and G. R. Kumar, Optics letters 29, 2662 (2004).

[30] A. Zigler, S. Eisenan, M. Botton, E. Nahum, E. Schleifer, A. Baspaly, I. Pomerantz, F. Abicht, J. Branzel, G. Priebe, S. Steinke, A. Andreev, M. Schnuerer, W. Sandner, D. Gordon, P. Sprangle, and K. W. D. Ledingham, Physical review letters 110, 215004 (2013).

[31] D. Margarone, O. Klimo, I. J. Kim, J. Prokůpek, J. Limpouch, T. M. Jeong, T. Mpcek, J. Pšikal, H. T. Kim, J Proška, K. H. Nam, L. Štolcova, I. W. Choi, S. K. Lee, J. H. Sung, T. J. Yu, and G. Korn, Physical review letters 109, 234801 (2012).

[32] M. A. Purvis, V. N. Shlyaptsev, R. Hollinger, C. Bargsten, A. Pukhov, A. Prieto, Y. Wang, B. M. Luther, L. Yin, S. Wang, and J. J. Rocca, Nature photonics 7, 796-800(2013).

[33] S. Jiang, L. L. Ji, H. Audesirk, K. M. George, J. Snyder, A. Krygier, P. Poole, C. Willis, R. Daskalova, E. Chowdhury, N. S. Lewis, D. W. Schumacher, A. Pukhov, R. R. Freeman, and K. U. Akli, Physical review letters 116, 085002 (2016).

[34] D. B. Zou, H. B. Zhuo, X. H. Yang, T. P. Yu, F. Q. Shao, and A. Pukhov, Physics of plasmas 22, 063103 (2015).

[35] I. A. Andriyash, R. Lehe, A. Lifschitz, C. Thaury, J. M. Rax, K. Krushelnick, and V. Malka, Nature communications 5, 4736 (2014).

[36] O. Klimo, J. Psikal, J. Limpouch, J. Proska, F. Novotny, T. Ceccotti, V. Floquet, and S. Kawata, New journal of physics 13, 053028 (2011).

[37] Y. Nodera, S. Kawata, N. Onuma, J. Limpouch, O. Klimo, and T. Kikuchi, Physical Review E 78, 046401(2008).

[38] L. L. Ji, J. Snyder, A. Pukhov, R. R. Freeman, and K. U. Akli, Scientific reports 6, 23256 (2016).

[39] J. Snyder, L. L. Ji, K. U. Akli, Physics of plasmas 23, 123122 (2016)

[40] Z. Gong, A. P. L. Robinson, X. Q. Yan, and A. V. Arefiev, arXiv preprint arXiv:1807.08075 (2018).

[41] D. Khaghani, M. Lobet, B. Borm, L. Burr, F. Gärtner, L. Gremillet, L. Movsesyan, and P. Neumayer, Scientific reports 7, 11366 (2017).

[42] B. Feng, L. L. Ji, B. F. Shen, X. S. Geng, Z. Guo, Q. Yu, T. J. Xu, and L. G. Zhang, Physics of plasmas 25, 103109 (2018).

[43] L. L. Ji, S. Jiang, A. Pukhov, R. R. Freeman, and K. U. Akli, High power laser science and engineering 5, (2017).

[44] L. X. Hu, T. P. Yu, Z. M. Sheng, J. Vieira, D. B. Zou, Y. Yin, P. McKenna, and F. Q. Shao, arXiv preprint arXiv:1804.08846 (2018).

[45] K. A. Ivanov, D. A. Gozhev, S. P. Rodichkina, S. V. Makarov, S. S. Makarov, M. A. Dubatkov, S. A. Pikuz, D. E. Presnov, A. A. Paskhalov, N. V. Eremin, and A. V. Brantov, Applied Physics B 123, 252 (2017).

[46] A. V. Arefiev, V. N. Khudik, and M. Schollmeier, Physics of plasmas 21, 033104 (2014).

[47] A. V. Arefiev, B. N. Breizman, M. Schollmeier, and V. N. Khudik, Physical review letters 108, 145004 (2012).

[48] P. L. Poole, C. Willis, R. L. Daskalova, K. M. George, S. Feister, S. Jiang, J. Snyder, J. Marketon, D. W. Schumacher, K. U. Akli, L. Van Woerkom, R. R. Freeman, and E. A. Chowdhury, Applied optics 55, 4713 (2016).

[49] K. A. Tanaka, T. Yabuuchi, T. Sato, R. Kodama, Y. Kitagawa, T. Takahashi, T. Ikeda, Y. Honda, and S. Okuda, Review of scientific instruments 76, 013507 (2005).

[50] N. Rabhi, K. Bohacek, D. Batani, G. Boutoux, J-E. Ducret, E. Guillaume, K. Jakubowska, C. Thaury, and I. Thfoin, Review of scientific instruments 87, 053306 (2016).

[51] Y. Sentoku, T. E. Cowan, A. Kemp, and H. Ruhl, Physics of plasmas 10, 2009 (2003)

[52] A. Pukhov, Journal of plasma physics 61, 425 (1999).

[53] A. V. Arefiev, G. E. Cochran, D. W. Schumacher, A. P. Robinson, and G. Chen, Physics of plasmas 22, 013103 (2015).

[54] A. P. L. Robinson, A. V. Arefiev, and V. N. Khudik, Physics of plasmas 22, 083114 (2015).

[55] J. Xu, B. Shen, X. Zhang, M. Wen, L. L. Ji, W. Wang, Y. Yu, and Y. Li, Physics of plasmas 17, 103108 (2010).